%% file: main.tex
\documentclass[submission,copyright,creativecommons]{eptcs}

\usepackage{cite}
\usepackage{amsmath,amssymb,amsfonts}
\usepackage{algorithmic}
\usepackage{graphicx}
\usepackage{svg}
\usepackage{textcomp}
\usepackage{xcolor}
\usepackage{tcolorbox}

\newcommand\blue[1]{\textcolor{blue}{#1}} 
\newcommand\cyan[1]{\textcolor{cyan}{#1}} 

\usepackage{braket} 

\usepackage{booktabs} 
\usepackage{multirow}
\usepackage{float}
\newfloat{Figure}{htbp}{lof}[section]
\usepackage{subfigure}
\usepackage{diagbox}
\usepackage{enumitem} 

\usepackage{url}
\usepackage{stmaryrd}
\usepackage{cleveref}
\usepackage{amsfonts}
\usepackage{makecell}

\input{macros} 

\title{On Refactoring Quantum Programs}

\author{Jianjun Zhao
\institute{Faculty of Information Science and Electrical Engineering}
\institute{Kyushu University, Japan}
\email{zhao@ait.kyushu-u.ac.jp}
}

\def\titlerunning{}
\def\authorrunning{}
\begin{document}
\maketitle

\begin{abstract}
Refactoring is a crucial technique for improving the efficiency and maintainability of software by restructuring its internal design while preserving its external behavior. While classical programs have benefited from various refactoring methods, the field of quantum programming lacks dedicated refactoring techniques. The distinct properties of quantum computing, such as quantum superposition, entanglement, and the no-cloning principle, necessitate specialized refactoring techniques. This paper bridges this gap by presenting a comprehensive set of refactorings specifically designed for quantum programs. Each refactoring is carefully designed and explained to ensure the effective restructuring of quantum programs. Additionally, we highlight the importance of tool support in automating the refactoring process for quantum programs. Although our study focuses on the quantum programming language Q\#, our approach is applicable to other quantum programming languages, offering a general solution for enhancing the maintainability and efficiency of quantum software.
\end{abstract}

\section{Introduction}
Refactoring is a disciplined process of improving the internal structure and reusability of a program without changing its external behavior~\cite{fowler1999refactoring,opdyke1992refactoring,griswold1992program}. It is a crucial practice that minimizes the introduction of bugs and enhances code quality. Developers continuously refactor code during its creation and throughout the software life cycle to improve the design. Refactoring tools enable software developers to safely and efficiently restructure their code, leading to enhanced quality, reliability, and maintainability.

Quantum programming is the process of designing and building executable quantum computer programs to achieve a particular computing result~\cite{ying2016foundations}. The field of quantum programming has gained significant attention due to its potential for solving complex computational problems. Various quantum programming approaches, such as Scaffold~\cite{abhari2012scaffold}, Qiskit~\cite{aleksandrowicz2019qiskit}, Q\#\cite{svore2018q}, ProjectQ\cite{haner2016high}, and Quipper~\cite{green2013quipper}, are available for writing quantum programs. With the maturing state of quantum programming research and the emergence of active research products, it is important for research on program refactorings to consider this new programming paradigm. Refactoring techniques tailored specifically for quantum programs can contribute to advancing the field and improving quantum software development.

The current research focus in quantum programming primarily revolves around problem analysis, software design, software testing and debugging, and implementation techniques~\cite{zhao2020quantum}. However, despite the importance of refactoring in improving software quality, the development of refactorings and tool support specifically tailored for quantum software has been largely overlooked. The unique properties of quantum computing, such as quantum superposition, entanglement, and the no-cloning principle, require specialized refactoring techniques to improve code quality in quantum software.

While refactoring has been extensively studied for classical software, no development of refactoring patterns for quantum software has been undertaken thus far. 
In this paper, we address this gap by presenting a comprehensive set of refactorings tailored specifically for quantum programs. Each refactoring is meticulously crafted and explained to facilitate the effective reconstruction of quantum programs. Furthermore, we emphasize the significance of tool support in automating the refactoring process for quantum programs. Although our study focuses on the quantum programming language Q\#, our approach holds generality and can be extended to other quantum programming languages to enhance the maintainability and efficiency of quantum software.

\begin{figure}[t]
\begin{center}
{\footnotesize
\begin{alltt}
     1   \blue{namespace} MyNamespace \{
     2       \blue{open} Microsoft.Quantum.Intrinsic;
     3
     4       \blue{operation} HelloWorld() : \blue{Unit} \{
     5           Message("Hello, quantum world!");
     6
     7           \blue{using} (qubit = \blue{Qubit}()) \{
     8              \cyan{// Apply Hadamard gate to create a superposition}
     9              H(qubit);
    10
    11              \cyan{// Apply CNOT gate for entanglement}
    12              \blue{using} (ancilla = \blue{Qubit}()) \{
    13                  CNOT(qubit, ancilla);
    14              \}
    15
    16              \cyan{// Measure qubits and display the result}
    17              \blue{let} result = M(qubit);
    18              \blue{let} entanglementResult = M(ancilla);
    19
    20              Message(\$"Measured qubit: \{result\}");
    21              Message(\$"Measured ancilla: \{entanglementResult\}");
    22
    23              \cyan{// Invoke the MultiplyByTwo function}
    24              \blue{let} multipliedResult = MultiplyByTwo(result);
    25              Message(\$"Multiplied result: \{multipliedResult\}");
    26           \}
    27       \}
    28
    29       \blue{function} MultiplyByTwo(x : Int) : \blue{Int} \{
    30          return 2 * x;
    31       \}
    32   \}
\end{alltt}
  \caption{\label{fig:sample} A sample Q\# program.}
}
\end{center}
\end{figure}

Given that quantum programming represents a new programming language paradigm diverging from classical programming, it is crucial to develop a systematic approach to support the refactoring of quantum software. By examining quantum refactoring ideas from multiple viewpoints and the independent development of quantum refactorings, we aim to gain a deeper understanding of the role of refactorings in the quantum programming domain and the significance of quality quantum software development. This paper serves as an initial report on our preliminary results in the field of quantum software refactoring.

The rest of the paper is organized as follows. Section~\ref{sec:QSharp} briefly introduces Q\#, a general quantum programming language developed by Microsoft. Section~\ref{sec:q-refactoring} presents a preliminary catalog of refactorings specifically designed for quantum programming. Section~\ref{sec:tool} discusses the tool support for refactoring quantum programs. Section \ref{sec:work} discusses related work, and concluding remarks are given in Section \ref{sec:conclusion}.

\section{Quantum Programming with Q\#}
\label{sec:QSharp}

We present our approach to quantum program refactoring using Q\# as our primary context. Q\# is a widely used quantum programming language for developing quantum algorithms and simulations~\cite{svore2018q}. Q\# enables developers to express quantum computations using a combination of classical control flow and quantum operations. In the following, We introduce some basic concepts in Q\#.

\vspace*{1mm}
\noindent
\textbf{Qubits.}
In Q\#, a qubit represents the fundamental unit of quantum information. It can be in a state of 0, 1, or a superposition of both. For example, we can create a qubit and set it to the superposition state using the Hadamard gate.

\vspace*{1mm}
\noindent
\textbf{Functions and Operations.}
Q\# supports both functions and operations. Functions are used to encapsulate reusable pieces of code that operate on classical data, while operations are used to define quantum algorithms and manipulate qubits. Functions can be called from within operations to perform classical computations in conjunction with quantum operations.

\vspace*{1mm}
\noindent
\textbf{Superposition and Entanglement.}
Superposition~\cite{nielsen2002quantum} is a fundamental concept in quantum computing, where a qubit can exist in multiple states simultaneously. By applying quantum gates, we can put a qubit in a superposition of 0 and 1. Entanglement~\cite{nielsen2002quantum} is another key concept that allows the correlation of multiple qubits. It enables the creation of quantum states that cannot be described independently for each qubit.

\vspace*{1mm}
\noindent
\textbf{Measurement.}
Measurement is a crucial operation in extracting classical information from quantum systems. In Q\#, the \texttt{M} operation is used to measure a qubit and return a classical bit as the result. In quantum computation, measurement results are probabilistic and follow a probability distribution. Therefore, repeating measurements on the same qubit may yield different outcomes.

\vspace*{1mm}
Figure \ref{fig:sample} shows a Q\# program that defines a Q\# operation called \textsf{\small HelloWorld} within the namespace \textsf{\small MyNamespace}. The \textsf{\small HelloWorld} operation starts with a message output, "Hello, quantum world!." It then creates a qubit using the \textsf{\small using} statement, which ensures proper resource management. The Hadamard gate \textsf{\small H} is applied to the qubit, putting it into a superposition state. Another qubit called \textsf{\small ancilla} is created using the \textsf{\small using} statement. The \textsf{\small CNOT} gate is applied between the original qubit and ancilla, creating entanglement. The \textsf{\small M} operation is used to measure both qubits and the measurement results are stored in variables \textsf{\small result} and \textsf{\small entanglementResult}, respectively. The measurement results are then displayed using the \textsf{\small Message} function. Finally, the \textsf{\small MultiplyByTwo} function is invoked with the result as the input, and the multiplied result is displayed. The \textsf{\small MultiplyByTwo} function simply multiplies the input x by 2 and returns the result.
This code provides a basic example of quantum computation in Q\#, showcasing concepts such as qubit creation, gates application, measurement, and function invocation.

\begin{figure}
\begin{center}
{\scriptsize
\begin{alltt}
\cyan{\textbf{// Before refactoring}}
\blue{namespace} MyNamespace \{

    \blue{operation} PerformQuantumSimulation(qubits : \blue{Qubit[]}, iterations : \blue{Int}) : \blue{Unit} \{
        \blue{for} (i \blue{in} 1..iterations) \{
            // Perform a series of quantum operations
            ApplyToEach(H, qubits);
            Controlled X(qubits[0..1], qubits[1]);
            Controlled X(qubits[0..1], qubits[0]);
            ApplyToEach(H, qubits);
            
            // Measure qubits and display the result
            \blue{let} measurements = MultiM(qubits);
            \blue{let} result = ResultArrayAsInt(measurements);
            Message(\$"Iteration {i}: Measurement result = {result}");
        \}
    \} 

    \blue{operation} Main() : \blue{Unit} \{
        \blue{using} (qubits = \blue{Qubit[2]}) \{
            PerformQuantumSimulation(qubits, 5);
        \}
    \}
\}

\cyan{\textbf{// After refactoring according to "split operation"}}
\blue{namespace} MyNamespace \{

    \blue{operation} PerformQuantumSimulation(qubits : \blue{Qubit[]}, iterations : \blue{Int}) : \blue{Unit} \{
        \blue{for} (i \blue{in} 1..iterations) \{
            PerformQuantumOperations(qubits);
            MeasureAndDisplayResult(qubits, i);
        \}
    \}

    \blue{operation} PerformQuantumOperations(qubits : \blue{Qubit[]}) : \blue{Unit} \{
        ApplyToEach(H, qubits);
        Controlled X(qubits[0..1], qubits[1]);
        Controlled X(qubits[0..1], qubits[0]);
        ApplyToEach(H, qubits);
    \}

    \blue{operation} MeasureAndDisplayResult(qubits : \blue{Qubit[]}, iteration : \blue{Int}) : \blue{Unit} \{
        \blue{let} measurements = MultiM(qubits);
        \blue{let} result = ResultArrayAsInt(measurements);
        Message(\$"Iteration {iteration}: Measurement result = {result}");
    \}

    \blue{operation} Main() : \blue{Unit} \{
        \blue{using} (qubits = \blue{Qubit[2]}) \{
            PerformQuantumSimulation(qubits, 5);
        \}
    \}
\}
\end{alltt}
  \caption{\label{fig:refactoring} An example for applying "Split Operation" refactoring.}
}
\end{center}
\end{figure}

\section{Refactorings for quantum programs}
\label{sec:q-refactoring}

We next present a catalog of refactorings for quantum programs that are essential for improving the design, readability, and maintainability of quantum programs. Unlike those refactorings for classical programs, these refactorings consider the unique properties of quantum computing. They are specially designed to optimize quantum programs for execution on real quantum hardware. By applying these refactorings, developers can simplify complex quantum algorithms, reduce the number of gates required for implementing computations, and minimize the risk of introducing errors or unwanted noise into the quantum system. These refactorings enhance the overall efficiency and reliability of quantum programs.

As we discussed in Section~\ref{sec:QSharp}, a Q\# program consists of two types of modules: \textit{functions} and \textit{operations}. Functions are used to encapsulate reusable pieces of code that operate on classical data, while operations are used to define quantum algorithms and manipulate qubits. For functions, we can use those classical refactorings introduced in~\cite{opdyke1992refactoring,fowler1999refactoring,mens2004survey,garrido2000software} to perform refactorings on them. In this paper, we only focus on the modules of operations in Q\# programs that are related to quantum operations and computations. 

\begin{table}
 \begin{center}
   \caption{\label{tab:q-refactoring} A Catalog of refactorings for quantum programs in Q\#}
\begingroup
\setlength{\tabcolsep}{4pt} 
\renewcommand{\arraystretch}{1.4} 
{\footnotesize
\begin{tabular}{|l|p{11.2cm}|}
\hline
\small{\bf Refactoring Name} & \small{\bf  Description} \\ 
\hline\hline

Add Parameter & Add new parameters to the definition of an operation\\
\hline

Consolidate Measurement & Consolidate multiple measurements in an operation into a single measurement. \\
\hline




Extract Function from Operation & Extract a portion of code within an operation into a separate function to promote code reuse and
modularity.\\
\hline

Extract Namespace & Move operations (functions) into a new namespace for better organization.
\\
\hline

Extract Operation & Extract a portion of code into a new operation to promote code reuse and modularity.
\\
\hline

Inline Function into Operation & Replace a function call in an operation with its actual code to simplify and streamline the operation.\\
\hline

Inline Operation &  Replace the invocation of an operation with its original code to eliminate unnecessary overhead and improve performance.\\
\hline

Introduce Classical Control & Add classical control statements, such as if-else conditions or loops, to control the execution of quantum gates within an operation.\\
\hline

Merge Gate & Combine adjacent gates into a single gate to optimize circuit execution.\\
\hline

Merge Operations & Combine multiple operations with similar functionality into a single operation to eliminate redundancy and improve code organization.
\\ 
\hline


Order Qubit & Rearranges the qubit order to optimize for specific quantum algorithms or hardware.\\
\hline

Parameterize Operation & Create one operation that uses a parameter for the different values for the problem that several operations do similar things but with different values contained in the operation body. \\
\hline



Remove Code Duplication & Identify and eliminate duplicated code within an operation to improve code efficiency and maintainability.\\
\hline

Remove Operation & Remove an unused operation that is no longer necessary to reduce clutter and improve code clarity.\\
\hline

Remove Parameter & Remove unused parameters from an operation that are no longer necessary to reduce clutter and improve code clarity.\\
\hline

Remove Variable & Remove unused variables from an operation that are no longer necessary to reduce clutter and improve code clarity.\\
\hline

Rename Variable & Change the name of a variable throughout and only in its scope. \\
\hline

Rename Parameter & Change the name of a parameter in an operation\\
\hline

Rename Operation & Rename operations with more meaningful and descriptive names to enhance code understanding and clarity.\\
\hline

Reorder Instructions & Reorder the sequence of instructions within an operation to optimize execution flow and minimize resource usage.\\
\hline

Reorder Parameters & Reorder the parameters in an operation definition and in all
calls to that operation.\\
\hline

Replace Gate & Replace specific gates with alternative gate sequences or equivalent gates that are more suitable for a specific quantum hardware platform.\\
\hline



Specialize Operation & Create specialized versions of an operation for specific use cases to optimize performance or accommodate specific constraints.\\
\hline

Split Operation & Divide a large operation into smaller, more focused operations to improve code modularity and maintainability.\\
\hline

Unroll Loop & Replaces a loop in an operation with its unrolled equivalent code, improving code execution efficiency.\\
\hline











\end{tabular}
}
\endgroup
\end{center}
\end{table}

\subsection{Motivating Example}

We present an example to show the necessity for performing refactoring for a quantum program. 
Figure \ref{fig:refactoring} presents a Q\# program containing two operations \textsf{\small Main} and \textsf{\small PerformQuantumSimulation}. 

In the original code, before refactoring, there is a single operation called \textsf{\small PerformQuantumSimulation} that performs a series of quantum operations within a loop. However, these operations are combined together using the \textsf{\small ApplyToEach} function, which makes it less clear what each individual operation does.
By applying the "split operation" refactoring, we can break down the combined operations into separate, more specific operations. This allows us to give each operation a meaningful name and make the code more self-explanatory.

In the refactored code, the original \textsf{\small PerformQuantumSimulation} operation has been split into two separate operations: \textsf{\small PerformQuantumOperations} and \textsf{\small  MeasureAndDisplayResult}. The \textsf{\small PerformQuantumOperations} operation focuses on executing the quantum gate operations, while the \textsf{\small MeasureAndDisplayResult} operation handles the measurement and display of results. 
By splitting the original operation, we improve code modularity and make it easier to understand and modify specific parts of the quantum simulation.
Each operation has a clear purpose and can be modified or reused independently. Additionally, the code becomes more modular, allowing for better organization and abstraction of the quantum operations.

\subsection{A Catalog of Refactorings for Quantum Programs}

Quantum programming introduces a unique set of refactorings specifically designed for quantum programs, setting them apart from classical refactorings. It is crucial to identify and present a comprehensive catalog of these quantum-specific refactorings to enable the effective and efficient development of quantum software. In Table \ref{tab:q-refactoring}, we present an initial list of the proposed refactorings for quantum programs. Each refactoring is listed in the first column, accompanied by a brief description of its purpose in the second column. It is important to note that this list is preliminary and will continue to expand as our research progresses. 


\section{Tool Support for Refactorings}
\label{sec:tool}

Tool support plays a crucial role in the successful application of refactoring techniques. In this section, we introduce our refactoring tool, named the \textit{QSharp Refactoring Tool (QRT)}, designed specifically to support the refactoring of Q\# programs. Unlike most existing refactoring tools that primarily operate on abstract syntax trees (ASTs) for classical code, QRT utilizes the \textit{program dependence graph (PDG)} as its underlying abstract data structure to represent quantum programs. This choice enables QRT to leverage the PDG for conducting refactoring operations on the programs. By adopting the PDG, QRT offers the advantage of automatically realizing a broader range of refactoring patterns compared to traditional AST-based tools.

\begin{figure}[t]
\centerline{\includegraphics[width=1.0\linewidth]{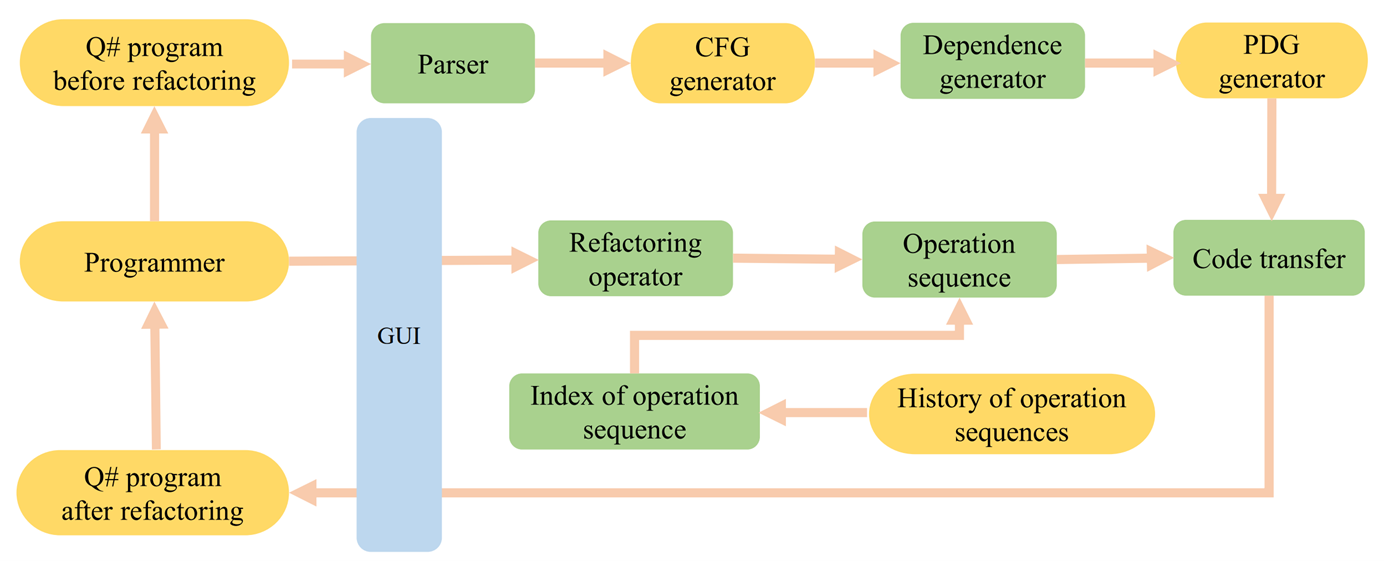}}
  \caption{\label{fig:QRT-structure} The structure of Q\# Refactoring Tool (QRT).}
\end{figure}

Figure \ref{fig:QRT-structure} illustrates the basic structure of QRT, which comprises two main components: the \textit{refactoring} and \textit{user interface components}. The refactoring component consists of a Q\# code parser, a dependence analyzer, a code transformer, a refactoring operator, and an operation sequence indexing module. These components work collaboratively to facilitate the refactoring process within QRT.

\section{Related work}
\label{sec:work}

Over the past decades, refactoring classical programs has emerged as a vibrant research area within the software engineering community, resulting in the development of numerous refactorings and associated support tools for classical programs \cite{fowler1999refactoring,opdyke1992refactoring}. 

However, due to the unique features of quantum programs, the majority of these refactorings cannot be directly applied to quantum programs due to their inability to handle the unique impact problem that arises when refactoring quantum software. To the best of our knowledge, this work represents the first attempt to address the challenge of refactoring quantum programs.

\section{Concluding Remarks}
\label{sec:conclusion}

This paper presents a comprehensive catalog of refactorings specifically tailored for quantum programs. Each refactoring is carefully designed and explained to facilitate effective and efficient refactoring of quantum code (Table~\ref{tab:q-refactoring}). We also emphasize the significance of automated tool support for refactoring quantum programs. Although our research focuses on the quantum programming language Q\#, our approach is also applicable to other quantum programming languages, enabling improved maintainability and efficiency of quantum software. 

Future work includes refining the refactoring catalog to incorporate more meaningful refactorings and developing dedicated refactoring tools to automate and streamline the process.

\bibliographystyle{plain}
\bibliography{qse-bibliography.bib}

\end{document}

%% file: macros.tex
\usepackage{times}
\usepackage{graphicx}
\usepackage{epsf}
\usepackage{verbatim}
\usepackage{url}
\usepackage{color}
\usepackage{alltt}

\newcommand{\Comment}[1]{}

\newcommand{\SmallSpace}{\vspace*{-1.4ex}}

